\documentclass[11pt]{article}
\usepackage{amssymb,bm,hyperref}

\newcommand{\di}[1]{\mathrm{d} #1}
\newcommand{\dR}{\mathbb{R}}
\newcommand{\dN}{\mathbb{N}}
\newcommand{\M}{\mathcal{M}}

\begin{document}

\noindent
\begin{center}

{\huge Local Thermal Equilibrium States and Unruh Detectors in Quantum Field
Theory} \\[25pt]
{\large \sc Jan Schlemmer}\\[7pt]
%
{ \it MPI for Mathematics in the Sciences, 
Inselstrasse 22, 04103 Leipzig, and \\
Inst. f{\"u}r Theoretische Physik, U. Leipzig, \
Postfach 100920, 04009 Leipzig \\
e-mail: Jan.Schlemmer@mis.mpg.de}
\end{center}

\indent

\begin{abstract}
In the framework of local thermodynamic equilibrium by
Buchholz, Ojima and Roos, a class $\mathcal{S}_x$ of observables,
whose members are supposed to model idealized measurements of 
thermal properties of given states at spacetime points $x$, plays 
a crucial role in determining and characterizing local equilibrium 
states in quantum field theory. Here it will be shown how elements
from this space can be reproduced by a specific model of the idealized 
measurements modeled by an Unruh-de Witt detector.
\end{abstract}



\section{Introduction}
In \cite{buojro} a method to identify states in relativistic quantum
field theory that allow locally a thermodynamic interpretation without
being in global thermodynamic equilibrium has been proposed. The basic 
procedure is the following: First one fixes at each spacetime 
point some (linear) space of observables, denoted by $\mathcal{S}_{x}$, 
which model idealized measurements of thermal properties in arbitrarily 
small spacetime regions. Next one chooses a set of reference states with 
known thermal properties (in the cases investigated so far this have been 
mixtures of global equilibrium states). A given state is then called 
\emph{(locally) thermal at a spacetime point $x$} if there exists a 
reference state such that the expectation values of all elements in 
$\mathcal{S}_x$ in the two states agree, i.e. as far as local measurements 
of (expectation values of) thermal parameters are concerned, the given state 
looks like the reference state. \\
If a state fulfills this condition of thermality at $x$ one can then
consistently assign to it the thermal parameters whose local measurement
are modeled by elements of $\mathcal{S}_x$ and among them the relations
(e.g. equations of state) established for the reference states also hold 
true. As the reference state may vary from point to point, this yields for 
states fulfilling the thermality condition in a region $\mathcal{O}$ an 
assignment of thermal parameters depending on $x \in \mathcal{O}$ (e.g. a 
time- and space-dependent temperature). For further details and applications 
of the formalism to models see \cite{buojro}, \cite{buchholz}, \cite{bahr} 
and \cite{busch}. \\
If the set of reference states is fixed (e.g. as mixtures of global
equilibrium states), there remains the question of which 
$\mathcal{S}_x$-spaces one should pick. In the models considered so far,
the $\mathcal{S}_x$-spaces have been chosen as the spaces generated by so 
called ``balanced derivatives'' of Wick-squares\footnote{for curved spacetimes
the ``general covariant Wick-square'' \cite{holwal},\cite{brufrever} can be 
used for a similar definition} 
, i.e. by elements
\begin{displaymath}
\lim_{\zeta \to 0} \partial_{\zeta}^{\bm{\mu}} \left(
\phi(x+\zeta)\phi(x-\zeta) - \omega_{\infty}\left(\phi(x+\zeta)\phi(x-\zeta)
\right) 1 \right)
\end{displaymath}
($\bm{\mu}$ a multiindex, $1$ the identity operator and $\omega_{\infty}$ 
the vacuum-state). Now
a justification of this choice of $S_x$-spaces is given in the models by
the fact that it allows for a local assignment of the interesting
thermodynamic properties to non-equilibrium-states in the above sense while
still allowing non-trivial local-equilibrium states (i.e. states where
the thermal parameters do vary in space and time). There is also a general 
discussion on the specification of a maximal space $\mathcal{T}_x$ of 
``pointlike'' observables that may be used to determine local thermal 
properties of states \cite{buojro}; the above set $\mathcal{S}_x$ should be 
a proper subset of $\mathcal{T}_x$. \\
As the concept of local thermodynamic equilibrium states under discussion
is based on the idea of idealized measurements at a point, one could hope 
that a more detailed description of such an idealized measurement in a model 
could give some additional insight into the choice of $\mathcal{S}_x$ spaces.
That this is indeed true  will be shown in this article. \\
As a model for the measurement process some kind of Unruh-de Witt detector
\cite{Unr},\cite{dW}, in this case a two-level system moving along
a given trajectory in spacetime and interacting with the
given quantum-field, is used. For large interaction-times and 
time-invariant states the (suitably normalized) probability for the 
transition of the detector system from its initial (ground- or excited
state) to its respective other state due to interactions with the
quantum field can be used to determine thermal properties of the quantum
field. Namely, by the principle of detailed balancing in a thermal
state one expects the transition rates from the ground- to the excited
and from the excited- to the ground-state to be related by a
Boltzmann factor $e^{-\beta E / \hbar}$ if the state of the field is
a thermal state at inverse temperature $\beta$ and $E$ is the
energy difference between the two levels of the detector system.
In fact, such a relation between the two transition rates for all two-level 
monopole detector ``at rest'' wrt. the thermal state\footnote{see section
5 below for the notion of being at rest relative to a thermal state} exactly 
corresponds to the KMS condition for this state \cite{takagi}.
More generally, the dependence of the transition 
probabilities on $E/\hbar$ corresponds to ``spectral properties'' of the 
field, with the principle of detailed balancing as a relation between the 
rates for $E$ and $-E$ as a special case.\\
Now in order to be able to proceed to measurements taking place in a
short time-interval, the idea is first not to look at the absolute
transition probabilities but rather at their differences to those in a common 
reference state (i.e. to ``remove the vacuum fluctuations'') and secondly to
consider instead of the transition probabilities as function of $E$ the 
sequence of moments of this function. When proceeding to arbitrarily
short measurement times (while increasing the interaction coupling suitably)
these moments will, in general, still diverge; however starting from the
zeroth moment which stays finite one can subtract from the higher moments
``perturbations'' by the lower moments in such a way that the resulting
(modified) moments all stay finite when sending the duration of measurement 
to zero. In this limit the modified moments are exactly what is measured
by the balanced derivatives ``in timelike direction'', i.e. by the
balanced derivatives with $\zeta$ tangential to the trajectory of the
detector. \\
Finally it will be shown that although these balanced derivatives do
not span the whole $\mathcal{S}_x$-spaces, their linear combinations
can be used to obtain the relevant thermodynamic observables that
so far appeared in the model of the massless, neutral Klein-Gordon field 
\cite{buchholz}.

\section{Specification of the model}
The quantum field considered is the massless neutral Klein-Gordon field
on Minkowski spacetime $(\dR^4,\eta)$, described by a star-algebra 
$\mathcal{A}$ generated by a unit element $1$ and elements $\phi(f)$ 
depending linearly on 
$f \in C_0^{\infty}(\dR^4)$ with the additional relations
\begin{itemize}
\item $\phi(f)^* = \phi(\overline f)$
\item $\phi(\Box f) = 0$
\item $[\phi(f),\phi(g)]= E(f,g) 1$ 
\end{itemize}
where 
\begin{displaymath}
E(f,g):=\int_{\dR^4} \int_{\dR^4} f(x) (G_{adv}-G_{ret})(x,x') g(x') \di{x}\di{x'}
\end{displaymath}
and $G_{adv}$ and $G_{ret}$ are the advanced and retarded fundamental
solutions of the Klein-Gordon equation, respectively. \\
States $\omega$ of the field are given by functionals on $\mathcal{A}$ which 
in addition fulfill for all $A \in \mathcal{A}$ and $f_1, \ldots f_n \in 
C_0^{\infty}(\dR^4)$,
\begin{itemize}
\item $\omega(A^* A) \geq 0$
\item $\omega(1) = 1$
\item $(f_1, \ldots, f_n) \mapsto \omega(\phi(f_1) \ldots \phi(f_n))$ is a 
distribution (depends continuously on the $f_i$).
\end{itemize}
There is a distinguished vacuum-state, denoted by $\omega_{\infty}$, and
furthermore the states considered here are Gaussian Hadamard states, i.e.
are determined by their two-point function $C_0^{\infty}(\dR^4) 
\times C_0^{\infty} (\dR^4) \ni (f,g) \mapsto \omega(\phi(f) \phi(g))$ 
which is such that $(f,g) \mapsto \omega(\phi(f) \phi(g)) -
\omega_{\infty}(\phi(f) \phi(g))$ is a regular distribution. \\
The physical picture of the measurements considered here is the following: An 
ensemble of quantum-mechanical detectors (two-level systems) moves along a 
(common, classical) trajectory $\gamma$, parametrized by proper time $\tau$, 
with each member initially in its ground-states. At some time the detectors 
are (smoothly) switched on, interact with the field for some time and are 
then switched off again. Finally the number of detectors in the excited state
is determined, which yields the transition probability for a single detector 
(which is of course the same as the expectation value of the transition rate 
times the interaction duration). \\
Mathematically, the free two-level detector system is described in the 
Heisenberg picture by a two-dimensional complex Hilbert-Space 
$(\mathcal{H}_D,<\cdot,\cdot>_D)$ spanned by the two orthonormal states 
$\psi_g$ and $\psi_e$ of the detector. These two states are assumed to be 
eigenstates with eigenvalues zero and $\epsilon$ of the detector-Hamiltonian 
$H_D$ and furthermore the existence of a (time-dependent) self-adjoint 
operator $\tau \mapsto M(\tau) := e^{i \tau H_{D}} M_0 e^{-i \tau H_{D}}$ such
that $\left \vert <\psi_e, M_0 \psi_g> \right \vert \neq 0$ is assumed. \\
For a given state $\omega$ of the quantum field, the
coupled detector-field system is described in the interaction picture in the
Hilbert-space $\mathcal{H}_{\omega} \otimes \mathcal{H}_{D}$ where 
$(\mathcal{H}_{\omega}, \pi_{\omega},\Omega_{\omega})$ denotes the GNS 
representation \cite[Chap. 4]{walqftcs} of $\mathcal{A}$ belonging to 
$\omega$. The initial state $\Phi$ of the coupled system is taken to be 
$\Omega_{\omega} \otimes \psi_g$ and the time evolution of this state 
is determined by 
\begin{displaymath}
i \partial_{\tau} \Phi(\tau) = H_{int}(\tau) \Phi(\tau) := 
\left[ \chi(\tau) \phi(\gamma(\tau)) \otimes M(\tau)\right] \Phi(\tau)
\end{displaymath}
where $\gamma: \dR \to \dR^4$ is the detector-trajectory as described above, 
$\chi \in \mathcal{S}(\dR)$ is the detector switching function (real-valued
and normalized by the requirement $\int_{\dR} \chi(\tau) \di{\tau} = 1$) and 
$\dR^4 \ni x \mapsto \phi(x)$ is related to $\pi_{\omega}(\phi(f))$  by 
$\pi_{\omega}(\phi(f)) = \int \phi(x) f(x) \di{x}$ in the sense of quadratic 
forms on (a subset of) $H_{\omega}$. \\
As one is interested in the transition probabilities of a (weakly) coupled
detector without back-reaction effects a perturbative calculation seems 
appropriate; for a rigorous justification of this approach in a related 
situation see \cite{biemerk}.
To first order perturbation-theory, the state of the detector-field system
at time $\tau$ is given by
\begin{displaymath}
\Phi(\tau) = \Omega_{\omega} \otimes \psi_g - i \int_{-\infty}^{\tau}
\chi(\tau') \ \left( \phi(\gamma(\tau')) \Omega_{\omega}\right) \otimes 
\left( M(\tau') \psi_g \right) \di{\tau'}
\end{displaymath}
For the probability of finding the detector in the 
excited state $\psi_e$ and the field in a state $\Psi_n$ at large times, one 
then has
\begin{eqnarray*}
\left \vert \langle \Psi_n \otimes \psi_e, 
\Phi(\infty) \rangle \right \vert^2 & = &  
\int_{\dR} \int_{\dR} \langle \chi(\tau') \phi(\gamma(\tau')) 
\Omega_{\omega} \otimes M(\tau') \psi_g, \Psi_n \otimes \psi_e \rangle 
\times \ldots \\ & \ldots & \times \langle
\Psi_n \otimes \psi_e, \chi(\tau'') \phi(\gamma(\tau''))\Omega_{\omega} 
\otimes M(\tau'') \psi_g \rangle \di{\tau'}\di{\tau''}
\end{eqnarray*}
and by summing over a complete set of $\Psi_n$ in $\mathcal{H}_{\omega}$
and using 
\begin{displaymath}
<\psi_e,M(\tau) \psi_g> 
= <\psi_e, e^{i \tau H} M_0 e^{-i \tau H} \psi_g>
= e^{i \epsilon \tau} <\psi_e, M_0 \psi_g>
\end{displaymath}
the probability of finding the detector in the excited and the field in any 
state is 
\begin{eqnarray*}
P_{\omega}(\epsilon) & = &  \left\vert <\psi_e,M_0 \psi_g> \right\vert^2 
\int_{\dR} \int_{\dR} \chi(\tau') \chi(\tau'') 
e^{-i \epsilon (\tau' - \tau'')} \times \ldots \\
& \ldots & \times <\phi(\gamma(\tau')) \Omega_{\omega},
\phi(\gamma(\tau'')) \Omega_{\omega} > \di{\tau'} \di{\tau''} \\
& = & m 
\int_{\dR} \int_{\dR} \chi(\tau') \chi(\tau'') 
e^{-i \epsilon (\tau' - \tau'')} \omega(\phi(\gamma(\tau'))\phi(\gamma(\tau''))) \di{\tau'} \di{\tau''}
\end{eqnarray*}
where the constant of proportionality $m:=\left\vert <\psi_e,M_0 \psi_g> 
\right\vert^2$ depends on details of the detector but not on the field 
configuration. \\
Now instead of comparing directly the transition probabilities
$P_{\omega}(\epsilon)$ and $P_{\omega_{\beta}}(\epsilon)$ in two states,
one can in principle compare their difference to the transition
probability $P_{\omega_{ref}}$ in a common reference state. Choosing
$\omega_{\infty}$ as $\omega_{ref}$ this amounts heuristically to
``removing the vacuum fluctuations'' and the resulting (difference in)
transition probability is then
\begin{eqnarray}
P^{ren}_{\omega}(\epsilon)  & = & m \int_{\dR} \int_{\dR}
e^{-i \epsilon s} 
F_{\omega}(\tau,s)
\chi \left( \tau + s/2 \right)
\chi \left( \tau - s/2 \right)
\di{s} \di{\tau}
\label{transprob} \\
F_{\omega} (\tau,s) & := & \omega \left(\phi\left(\gamma\left(\tau+s/2\right)
\right) \phi \left(\gamma\left(\tau-s/2\right)\right)\right) - 
\ldots \nonumber \\
& & \ldots - \omega_{\infty} \left(\phi\left(\gamma\left(\tau+s/2\right)
\right) \phi \left(\gamma\left(\tau-s/2\right)\right)\right) \nonumber
\end{eqnarray}
Assuming $\omega$ to be a Hadamard state, the integrand is smooth
and compactly supported and therefore $\epsilon \mapsto P^{ren}_{\omega}
(\epsilon)$ is a rapidly decreasing, smooth function. Thus all moments
of this function are defined and we turn to the analysis of these
moments in order to obtain a means for local investigation of
states.

\section{Convolution and moments}
As already mentioned above, in the case of global equilibrium and
for $\chi$ approaching a constant function, the dependence
of the transition probabilities on $\epsilon$ approaches a function that
gives information about the thermal properties of the state under 
consideration. Disregarding for a moment the $\tau$-integration
in (\ref{transprob}), more rapid switching of the detector can be
seen to disturb this function by convolution with a function that
becomes wide as $\chi$ becomes narrow, as is of course to be expected
due to time-energy uncertainty. There is however a way to get
around this, if one is only interested in the moments of this function 
and knows the moments of $\chi$. \\
To see this, consider two rapidly decaying functions 
$f, h \in \mathcal{S}(\dR)$. Denote the $k$-th 
moment of a function $f\in\mathcal{S}(\dR)$ by
\begin{displaymath}
\M^k[f]:=\int_{\dR}t^k f(t) \di{t}
\end{displaymath}
and assume $\M^0[h]=1$ (i.e. the integral over $h$ is one). Then by a direct 
computation one has for the convolution $f*h$:
\begin{equation}
\M^k[f*h] = \sum_{j=0}^k {k \choose j} \M^j[f] \M^{k-j}[h]
\label{momeqn}
\end{equation}  
For the special case $k=0$ this relation gives 
$\M^0[f * h]= \M[f]$, so the zeroth moment of $f$ is always known once
the zeroth moment of $f * h$ is known. Noting that on the rhs of
equation (\ref{momeqn}) the $k$-th moment of $f$ is multiplied by one
and the further terms involve only lower moments of $f$, one can therefore 
determine the moments of $f$ from those of $f*h$ and $h$ by
the recursion
\begin{equation}
\M^k[f] = \M^k[f*h]- \sum_{j=0}^{k-1} {k \choose j} \M^j[f] \M^{k-j}[h]
\label{receqn}
\end{equation}
starting from $\M^0[f]$.

\section{Modified moments of transition rates and elements of $\mathcal{S}_x$}
Now returning to (\ref{transprob}), for general $\chi$ the application of 
the idea of the preceding section to the transition-probability is 
complicated by the $\tau$-integration. This problem can however be overcome 
by choosing $\chi$ to be a Gaussian of width $\sigma$ centered at $\tau_0$:
\begin{displaymath}
\chi_{\sigma,\tau_0}: \tau \mapsto \frac{1}{\sigma \sqrt{2 \pi}}
e^{-\frac{(\tau-\tau_0)^2}{2 \sigma^2}} 
\end{displaymath}
Then $\chi_{\sigma,\tau_0} \left( \tau + s/2 
\right) \chi_{\sigma,\tau_0} \left( \tau - s /2 \right)$ factorizes into
$\chi_{\sigma/\sqrt{2},\tau_0} (\tau) \chi_{\sqrt{2} \sigma,0} (s)$
and the moments of $\epsilon \mapsto P^{ren}_{\omega}(\epsilon)$ are:
\begin{displaymath}
\M^k[P^{ren}_{\omega}]  = m \int_{\dR} \int_{\dR} \int_{\dR} 
F_{\omega}(\tau,s) \chi_{\sigma/ \sqrt{2},\tau_0} ( \tau ) \di{\tau} \
e^{-i \epsilon s} \chi_{\sqrt{2} \sigma,0} (s) 
\di{s} \epsilon^k \di{\epsilon}
\end{displaymath}
Now identifying the Fourier transform of $s \mapsto \sqrt{2 \pi}
\int_{\dR} F_{\omega}(\tau,s) \chi_{\sigma / \sqrt{2}} (\tau) 
\di{\tau}$ with $f$ and the Fourier transform of 
$\sqrt{2 \pi} \chi_{\sqrt{2} \sigma}$ with $h$, this is the situation of the 
preceeding section and by the recursive procedure described there, one can 
obtain the $k$-th moments of $f$ which will be called 
$\mathcal{P}^k_{\omega}$. Explicitly they are given by
\begin{displaymath}
\mathcal{P}^k_{\omega} = m \int_{\dR} \int_{\dR} F_{\omega}(\tau,s) 
\chi_{\sigma / \sqrt{2},\tau_0}(\tau) \di{\tau} e^{-i \epsilon s} \di{s}
\epsilon^k \di{\epsilon}
\end{displaymath}
By partial integration using the decay-properties of $\chi_{\sigma,\tau_0}$ 
and the regularity of $F_{\omega}$ this can be expressed as
\begin{displaymath}
\mathcal{P}^k_{\omega} = m \int_{\dR} (-i \partial_{s})^k 
F_{\omega}(\tau,s) \vert_{s = 0} 
\chi_{\sigma / \sqrt{2},\tau_0}(\tau) \di{\tau} 
\end{displaymath}
For this expression one can proceed to the limit $\sigma \to 0$,
which shows that the $\mathcal{P}^k_{\omega}$ are objects which
can be measured over arbitrarily short time-intervals. In the limit
one has
\begin{eqnarray}
\lim_{\sigma \to 0} \mathcal{P}^k_{\omega} & = & m (-i \partial_{s})^k
F_{\omega}(\tau_0, s) \vert_{s = 0} \nonumber \\
& = & m (-i/2)^k \partial_{s}^k
(\omega-\omega_{\infty}) \left( \phi(\gamma(\tau_0+s))
\phi(\gamma(\tau_0-s)) \right) \vert_{s=0} 
\label{reseqn}
\end{eqnarray} 
Defining for a (sufficiently differentiable) function 
$f: \dR^4 \times \dR^4 \to \dR$ and $u_1, \ldots, u_n \in \dR^4$ the 
\textit{balanced derivative} $\eth^{(u_1, \ldots, u_n)} f(x)$ by
\begin{equation}
\eth^{(u_1, \ldots, u_n)} f(x) = \partial_{t_1} \ldots 
\partial_{t_n} f\Big(x+\sum_{i=1}^n t_i u_i \ , \
x-\sum_{j=1}^n t_j u_j\Big) \vert_{t_1=\ldots=t_n=0}
\label{balderdef}
\end{equation} 
one can rewrite (\ref{reseqn}) for inertial detectors with $\gamma(\tau_0)=x$
and $\dot \gamma (\tau_0) = u$ as
\begin{eqnarray*}
\lim_{\sigma \to 0} \mathcal{P}_{\omega}^k & = & m (-i/2)^k \ 
\eth^{(u, \ldots, u)}  w(x) \\
w(x,y) & := & \omega(\phi(x)\phi(y)) - \omega_{\infty}(\phi(x)\phi(y))
\quad \mathrm{in} \ \mathrm{the} \ \mathrm{sense} \ \mathrm{of} \ 
\mathrm{distrib.}
\end{eqnarray*}
The definition (\ref{balderdef}) slightly generalizes the definition for 
$\eth^{(\mu_1, \ldots, \mu_n)}$ from \cite{buojro} which is recovered
by choosing $u_i = \eta^{\mu_i \mu_i} e_{\mu_i}$ with $e_0, \ldots, e_3$ the 
basis vectors of $\dR^4$, so the expectation values of 
elements $m (-i/2)^k \ \eth^{(0, \ldots, 0)} : \phi^2 :$ from the 
$S_x$-spaces in the state $\omega$ used there can be interpreted as 
measurements of $\mathcal{P}^{k}_{\omega}$ in the limit of arbitrarily 
short interaction of detector and field. A prominent example from this class 
is the Wick-square itself, which gives the expectation-value of the local 
temperature squared.

\section{Moving detectors and comparison with the full $\mathcal{S}_x$-space}
As established in the last section, the measurement of balanced
derivatives $\eth^{(u, \ldots, u)} : \phi^2 (x):$ can be described by
a limiting process involving measurements carried out on an ensemble of
detectors moving through the spacetime point $x$ with a four-velocity $u$.
These balanced derivatives however only generate a subset of the 
$\mathcal{S}_x$-spaces and important (local thermal)
observables like the (thermal) stress-energy tensor are not among them. \\
Now global equilibrium states on Minkowski spacetime are not invariant under 
the full Poincar\'e group but single out a set of inertial frames that only
differ by rotations and translations and physically correspond to the
observers being at rest with respect to the ``gas'' described by the thermal 
state.
As thermality properties like the principle of detailed balancing
hold only for those systems coupled to the field which are at rest in these 
inertial frames, for the investigation of an equilibrium state with an unknown
associated rest frame one should use not one detector, but 
detectors with all possible velocities smaller than the velocity of light 
relative to a given one. The detector behaving according to the principle of 
detailed balancing then indicates the rest frame of the given state, and 
starting from this one can then check whether the readings of the other 
detectors are compatible with the interpretation of being in relative motion 
to a thermal state. \\
Whereas for a global equilibrium state, whose rest-frame is usually 
known a priori this discussion might sound rather odd, 
the hydrodynamical description of gases by a velocity field varying in
space and time can be rephrased as the statement that at each point the
state looks like a thermal state with reference frames at different 
points being in relative motion to each other. As this dependence of the
frames on space-time is not known a priori but rather one of the informations
an LTE-formalism should yield, it seems therefore sensible to not
just consider one detector with a worldline passing through a spacetime point
$x$ but the set of all detectors passing through it. \\
By the above procedure, one then obtains as local thermal observables the
balanced derivatives $\eth^{(u, \ldots, u)} :\phi^2:(x)$ for \emph{all} 
time unit vectors $u$. By multiplying with scalars, the requirement of 
normalization for $u$ can be dropped, and forming linear combinations one 
furthermore has
\begin{eqnarray*}
\frac{1}{4} \eth^{(e_j, e_j)} :\phi^2: & = & 
\eth^{\left(e_0 + \frac{1}{2} e_j, e_0 + \frac{1}{2} e_j \right)} :\phi^2: 
+ \eth^{\left(e_0 - \frac{1}{2} e_j, e_0 - \frac{1}{2} e_j\right)} 
:\phi^2: - \ldots \\
 & \ldots & - \eth^{(e_0, e_0)} : \phi^2 : \\
2 \eth^{(e_0, e_j)} : \phi^2 : & = &
\eth^{\left(e_0 + \frac{1}{2} e_j, e_0 + \frac{1}{2} e_j\right)} : \phi^2 :
- \eth^{\left(e_0 - \frac{1}{2} e_j, e_0 - \frac{1}{2} e_j\right)} 
: \phi^2 : \\
\eth^{(e_j, e_k)} :\phi^2 : & = & \eth^{\left(e_0 + \frac{1}{2} e_j 
+ \frac{1}{2} e_k,e_0 + \frac{1}{2} e_j + \frac{1}{2} e_k \right)} : \phi^2 : 
- \ldots \\ & \ldots & - \eth^{\left(e_0 + \frac{1}{2} e_j - \frac{1}{2} e_k,
e_0 + \frac{1}{2} e_j - \frac{1}{2} e_k \right)} : \phi^2 : - \ldots \\ 
& \ldots & - \eth^{\left(e_0 + \frac{1}{2} e_k, e_0 + \frac{1}{2} e_k \right)}
: \phi^2: + \eth^{\left(e_0 - \frac{1}{2} e_k, e_0 - \frac{1}{2} e_k\right)}
: \phi^2:
\end{eqnarray*}
where $k\neq j, k,j=1,2,3$ and $e_0, \ldots, e_3$ are the basis-vectors of 
$\dR^4$ which shows that balanced derivatives $\eth^{(\mu,\nu)} :\phi^2:$ for
$\mu, \nu = 0, \ldots, 4$ can be expressed as linear combinations of balanced
derivatives $\eth^{(u,u)} :\phi^2:$ with different, timelike $u$ (so that e.g.
the expectation values of the thermal stress-energy tensor at $x$ can be
determined as a linear combination of measurement results of detectors moving
through $x$ with different velocities). \\
In the models considered so far it was also possible to locally determine 
the entropy current and the particle density by introducing an approximation
process in the $\mathcal{S}_x$-spaces adapted to the notion of local 
thermality used. For a detailed description and motivation see \cite{buojro};
at a technical level this approximation criterion boils down to the following:
Given a function $V^+ \ni \beta \mapsto \Xi(\beta) \in \dR$ describing the 
(known) dependence  of a macroscopic observable $\Xi$ on the (parameters 
$\beta$ labeling) global equilibrium states, try to approximate $\Xi$ in the 
seminorms
\begin{equation}
\sup_{\beta \in B} \left\vert \Xi(\beta) \right\vert =: \left\Vert \Xi
\right\Vert_{B}
\label{normdef} 
\end{equation}
($B$ a bounded set contained inside the open forward lightcone $V^+$)
by the ``thermal functions'' of the elements in $\mathcal{S}_x$. The
thermal function of an element from $\mathcal{S}_x$ is defined as the mapping
that associates to $\beta$ the expectation value of this element in the
global-equilibrium state belonging to $\beta$; those associated to
$\eth^{\bm{\mu}} :\phi^2:(x)$ will be denoted by $\Theta^{\bm{\mu}}$. \\
Investigating the global equilibrium states for the neutral, massless 
Klein-Gordon field (labeled by a timelike four-vector $\beta$ already used 
above), the functions belonging to the particle density and the entropy
current are found to be, respectively, a solution of the wave-equation and
a gradient of such a solution (wrt. the parameter $\beta$) \cite{buojro},
\cite{buchholz}. That such functions can indeed be approximated in the above
sense is established in \cite{buchholz}. The approximation given there
only needs thermal functions $l_{\bm{\mu}} \Theta^{\bm{\mu}}$ corresponding 
to $\eth^{(l, \ldots, l)}: \phi^2 :$ with $l$ lightlike\footnote{In this 
section $\bm{\mu}$ is always assumed to be a multi-index and the Einstein 
summation convention is used}, whereas here an 
approximation with thermal functions $u_{\bm{\mu}} \Theta^{\bm{\mu}}$ of 
elements $\eth^{(u, \ldots, u)} :\phi^2 :$ with $u$ timelike is desired. \\
Relying on these results, it is thus sufficient to
show that the functions $l_{\bm{\mu}} \Theta^{\bm{\mu}}$ for
given, lightlike $l$ can be approximated in the seminorms (\ref{normdef}) 
by functions $u_{\bm{\mu}} \Theta^{\bm{\mu}}$ with timelike vectors $u$. \\
For the massless, neutral Klein-Gordon field, the thermal functions
$u_{\bm{\mu}} \Theta^{\bm{\mu}}$ of elements $\eth^{(u, \ldots, u)} :\phi^2:$
are proportional to \cite{buojro}
\begin{displaymath}
u_{\bm{\mu}} \Theta^{\bm{\mu}} = \left( u^{\kappa} \frac{\partial}
{\partial \beta_{\kappa}} \right)^m \ \frac{1}{\beta^2}
\end{displaymath}
for $m:=\vert \bm{\mu} \vert$ even, and zero otherwise. This $m$-fold
directional derivative is given by a sum starting with the
term $m! 2^m (u \beta)^m (\beta^2)^{-m-1}$ and further terms proportional
to $(u^2)^k (u \beta)^l (\beta^2)^{-n} (k,n\in\dN\setminus\{0\}, l\in\dN)$ 
that vanish for $u$ lightlike. Now for given lightlike $l$ choose 
$u=l+\delta e_0$. The first term in the sum is then itself a sum
of $m! 2^m (l \beta)^m (\beta^2)^{-m-1}$ and functions bounded on $B$ times
powers of $\delta$ and the remaining summands also give functions bounded
on $B$ multiplied by powers of $\delta$ greater than zero (because of
the terms $(u^2)^k, k\in\dN\setminus\{0\}$). For given $B$,$l$ and $\epsilon$ 
one can therefore choose $\delta$ such that 
$(l+\delta e_0)_{\bm{\mu}} \Theta^{\bm{\mu}}$ approximates
$l_{\bm{\mu}} \Theta^{\bm{\mu}}$ in the $\Vert \cdot \Vert_B$-norm up to
an error $\epsilon$. \\
Summarizing, for the model of the massless, neutral Klein-Gordon field the
relevant local thermodynamic observables can all be obtained by
linear combinations and limiting processes involving the balanced
derivatives from section four, whose operational significance has been
established. \\
As far as the case of a massive Klein-Gordon field is concerned, the 
results up to and including the determination of the thermal stress-energy
tensor using detectors in relative motion do not depend on the assumption
of a vanishing mass; the question whether one can still approximate local 
observables for the particle density and the entropy current in the case of
a massive field is however more difficult to answer, as in this situation 
one has on the one hand to do the approximation in a different topology 
(for a definition and reasons for its use see \cite{huebner}) and on the 
other hand the thermal functions of the generators of $\mathcal{S}_x$ are 
more complicated.

\section{Conclusion}
Starting from an Unruh-de Witt detector model, it was shown that some
modified moments characterizing the dependence of the transition rate
of this detector on the separation of its levels survive the limit of 
going to arbitrarily short detector-field interactions (rescaling the 
strength of the interaction in the natural way). The value of these
moments obtained in the limit coincides for geodesic detectors in
Minkowski spacetime with the expectation values of the balanced
derivatives defined in the context of non-equilibrium thermodynamics.
Like other observables defining properties of quantum fields at a point
(for example the stress-energy tensor), the objects $\mathcal{P}^{k}_{\omega}$
have to be carefully defined in order to remain meaningful in the 
point-limit. Here this definition involves
\begin{itemize}
\item Looking at differences between transition probabilities in
different states instead of looking at the transition probabilities itself. 
\item ``Decoupling'' time and frequency domain by choosing Gaussian
switching functions, choosing moments of the transition probability and 
removing the perturbing effects of lower on higher moments by the
re\-cur\-sion-relation (\ref{receqn}).
\end{itemize}
The last point bears some similarity to a renormalization-group procedure:
When going to higher moments, the high-frequency (small scale) 
properties of the field are emphasized and at the same time the 
low-frequency part is discarded in order to be able to proceed to a 
pointlike limit.

\section*{Acknowledgements} 
I would like to thank Rainer Verch for valuable discussions
about this topic and help in formulating this article and the International 
Max Planck Research School for financial support.

\end{document}